\documentclass[10pt,a4paper,twocolumn,american,aps, amsmath, prb]{revtex4}

\usepackage[T1]{fontenc}
\usepackage[latin9]{inputenc}
\usepackage{color}
\usepackage{babel}
\usepackage{bm}
\usepackage{amsmath}
\usepackage{graphicx}
\usepackage{esint}
\usepackage[unicode=true,pdfusetitle,
 bookmarks=true,bookmarksnumbered=false,bookmarksopen=false,
 breaklinks=false,pdfborder={0 0 1},backref=false,colorlinks=false]
 {hyperref}
\usepackage{breakurl}

\makeatletter

\@ifundefined{textcolor}{}
{%
 \definecolor{BLACK}{gray}{0}
 \definecolor{WHITE}{gray}{1}
 \definecolor{RED}{rgb}{1,0,0}
 \definecolor{GREEN}{rgb}{0,1,0}
 \definecolor{BLUE}{rgb}{0,0,1}
 \definecolor{CYAN}{cmyk}{1,0,0,0}
 \definecolor{MAGENTA}{cmyk}{0,1,0,0}
 \definecolor{YELLOW}{cmyk}{0,0,1,0}
}

\PassOptionsToPackage{normalem}{ulem}
\usepackage{ulem}
\let\cite@rig\cite
\newcommand{\b@xcite}[2][\%]{\def\def@pt{\%}\def\pas@pt{#1}
  \mbox{\ifx\def@pt\pas@pt\cite@rig{#2}\else\cite@rig[#1]{#2}\fi}}
\renewcommand{\underbar}[1]{{\let\cite\b@xcite\uline{#1}}}

\AtBeginDocument{
  
}

\makeatother

\begin{document}

\title{AC susceptibility of an assembly of nanomagnets: combined effects
of surface anisotropy and dipolar interactions}

\author{F. Vernay,}

\email{francois.vernay@univ-perp.fr}

\selectlanguage{american}%

\author{Z. Sabsabi, and H. Kachkachi}

\email{hamid.kachkachi@univ-perp.fr}

\selectlanguage{american}%

\affiliation{Laboratoire PROMES-CNRS (UPR-8521) \& Université de Perpignan Via
Domitia, Rambla de la thermodynamique, Tecnosud, 66100 Perpignan,
FRANCE.}

\date{\today}

\begin{abstract}
We compute the AC susceptibility of a weakly dipolar-interacting monodisperse
assembly of magnetic nanoclusters with oriented anisotropy. For this
purpose we first compute the relaxation rate in a longitudinal magnetic
field of a single nanomagnet taking account of both dipolar interactions
in the case of dilute assemblies and surface anisotropy. We then study
the behavior of the real and imaginary components of the AC susceptibility
as functions of temperature, frequency, surface anisotropy and inter-particle
interactions. We find that the surface anisotropy induces an upward
shift of the temperature at the maximum of the AC susceptibility components
and that its effects may be tuned so as to screen out the effects
of interactions. The phenomenological Vogel-Fulcher law for the effect
of dipolar interaction on the relaxation rate is revisited within
our formalism and a semi-analytical expression is given for the effective
temperature is given\emph{ }in terms of \emph{inter alia }the applied
field, surface anisotropy and dipolar interaction.
\end{abstract}
\maketitle

\section{Introduction}

The dynamics of magnetic systems in the form of nanoclusters (nanoparticles
or nanomagnets) assemblies is a rather challenging issue from the
standpoint of fundamental physics as it requires a simultaneous investigation
of both long-range inter-cluster interactions and the intricacies
of inhomogeneous magnetism taking place inside the clusters. Even
for the equilibrium properties, the problem is of a tremendous difficulty
especially if one tries to take account of the internal structure
of the cluster by regarding it as a many spin system. In fact, only
advanced numerical approaches may offer a way out, though with a limited
success inasmuch as one considers the effect of surface anisotropy
and its interplay with the inter-cluster dipolar interactions. Recently,
this issue has been tackled \citep{margarisetal12prb,sabsabietal13prb}
to some extent by representing each nanocluster by an effective macroscopic
model \citep{garkac03prl,kacbon06prb,kachkachi07j3m,yanesetal07prb}
with an energy potential whose coefficients are functions of the cluster's
characteristics (size, shape, lattice crystal, spin-spin interactions).
It was shown that the magnetic properties of an assembly may be improved
by a tailored variation of the assembly parameters, such as its concentration
and geometry, and the clusters intrinsic characteristics such as the
size and shape. In this work, we investigate the joint effect of inter-cluster
interactions and surface anisotropy on the dynamic behavior of the
assembly, in the case of low concentration and not too strong surface
effects. For this we study the AC susceptibility with a variable measuring
frequency.

AC susceptibility of an assembly of magnetic nanoclusters has been
studied by many authors during the last decades, experimentalists
and theorists, by varying the applied magnetic field, temperature
and frequency.\citep{gitetal74prb,doretal97acp,luisetal04jpcm,tronc03jmmm,jongar01epl,bergor01jpcm,raishl94acp,svedlindhetal97j3m,raiste97prb,anderssonetal97j3m,garlaz98prb,spinu98phd,doretal99jmmm,jonetal01prb,raiste02prb,troetal03jmmm,brinisetal_JAP2014}
These studies have greatly contributed to improve our understanding
of the superparamagnetic behavior of such systems and to provide estimates
of their physical parameters. In particular, the size study \citep{troetal03jmmm}
of AC susceptibility, together with Mössbauer spectroscopy, of diluted
and concentrated assemblies of maghemite nanoclusters dispersed in
polymer, has revealed the important role of surface effects. On the
theoretical side, it is the first time that the joint effects of inter-cluster
interactions and surface anisotropy on the AC susceptibility are considered
in a single study. 

According to Debye's model applied to assemblies of magnetic nanoclusters,\citep{gitetal74prb}
the AC susceptibility is given by $\chi\left(\omega\right)=\chi_{\mathrm{eq}}/\left(1+i\omega\Gamma^{-1}\right)$,
where $\chi_{\mathrm{eq}}$ is the static or equilibrium susceptibility,
$\omega$ the frequency and $\Gamma$ the clusters relaxation rate
(inverse of relaxation time). This model describes the absorption
by a single mode of the electromagnetic energy provided by the applied
field. The dynamics of this mode is rather slow and characterized
by the longitudinal relaxation time $\tau=\Gamma^{-1}$ corresponding
to the population inversion from the blocked state to the superparamagnetic
state. This transition corresponds on average to the crossing by each
cluster's magnetic moment of its energy barrier. Therefore, in order
to compute the AC susceptibility, one has to compute the longitudinal
relaxation rate of a nanocluster in the assembly (described by an
effective model) in a magnetic field.

The paper is organized as follows: Section II is devoted to the presentation
of the model and the statement of the problem. This Section closes
with a brief summary of the results for the equilibrium susceptibility
obtained in Ref. \onlinecite{sabsabietal13prb} as a function of the
applied field, temperature, surface anisotropy and including the contribution
of long-range dipolar interaction. The formulas for the AC susceptibility
are then derived in Section III: we first describe in details the
evaluation of single nanocluster's relaxation rate $\Gamma$ with
both a uniaxial and a cubic anisotropy representing the surface effects;
by using Debye's model the semi-analytical form of the AC susceptibility
is then given at the end of the Section. In Section IV we deal with
the main focus of the present work, namely the study of the effect
of surface anisotropy on the AC susceptibility and its competition
with dipolar inter-particle interactions. The paper ends with a discussion
of the Vogel-Fulcher law and concluding remarks.

\section{Energy and equilibrium susceptibility}

\subsection{Nanoparticle assembly}

We consider a monodisperse and textured assembly of $\mathcal{N}$
ferromagnetic nanoclusters each carrying a magnetic moment $\bm{m}_{i}=m_{i}{\bf s}_{i},\, i=1,\cdots,{\cal N}$
of magnitude $m$ and direction ${\bf s}_{i}$, with $\vert{\bf s}_{i}\vert=1$.
Each magnetic moment has a uniaxial easy axis ${\bf e}$ aligned along
the $z$ direction. The energy of a magnetic moment $\bm{m}_{i}$
interacting with all the other magnetic moments within the assembly,
in a magnetic field $\bm{H}=H\bm{e}_{h}$, reads (after multiplying
by $-\beta=-1/k_{B}T$) 
\begin{equation}
\mathcal{E}_{i}={\cal E}_{i}^{(0)}+{\cal E}_{i}^{\mathrm{DDI}},\label{eq:DDIAssemblyEnergy}
\end{equation}
where the first contribution ${\cal E}_{i}^{(0)}=x_{i}{\bf s}_{i}\cdot{\bf e}_{h}+\mathcal{A}\left({\bf s}_{i}\right)$
is the energy of the free nanocluster at site $i$, comprising the
Zeeman energy and the anisotropy contributions from the core and surface.
$\mathcal{A}\left({\bf s}_{i}\right)$ is a function that depends
on the anisotropy model and is given by 
\begin{equation}
\mathcal{A}(\mathbf{s}_{i})=\left\{ \begin{array}{ll}
\sigma_{i}\left(\mathbf{s}_{i}\cdot\mathbf{e}_{i}\right)^{2}, & \ \mathrm{OSP},\\
\\
\sigma_{i}\left[\left(\mathbf{s}_{i}\cdot\mathbf{e}_{i}\right)^{2}-\frac{\zeta}{2}\left(s_{i,x}^{4}+s_{i,y}^{4}+s_{i,z}^{4}\right)\right], & \ \mathrm{EOPS}.
\end{array}\right.\label{eq:EnergyFreePart}
\end{equation}

OSP and EOSP stand respectively for One-spin problem and Effective
One-spin problem which are macroscopic models used for representing
the magnetic state of the nanocluster.\citep{sabsabietal13prb} In
the present case, \textcolor{black}{we restrict ourselves to the situation
where the uniaxial anisotropy axis is aligned along the $z$ direction,
}\textcolor{black}{\emph{i.e.}}\textcolor{black}{{} with a common axis
with the cubic anisotropy. This assumption makes the analytical calculations
somewhat simpler and the physical interpretation more transparent,
but it does not represent a significant discrepancy with regard to
the real situation. Indeed, the uniaxial anisotropy considered in
Eq. (\ref{eq:EnergyFreePart}) is in fact an effective anisotropy
that takes account of both the magneto-crystalline and shape anisotropy.
In typical nanoparticle assemblies this effective anisotropy is rather
strong, especially for elongated nanoparticles. As such a small tilting
of the cubic anisotropy with respect to the axis of the effective
uniaxial anisotropy should not change the results in a significant
way. For a more general situation with an arbitrary orientation of
the cubic anisotropy axes with respect to the uniaxial anisotropy
axis, one can write the cubic contribution in a different reference
frame $\left(x',y',z'\right)$ and then introduce in Eq. (\ref{eq:EnergyFreePart})
a rotation matrix such that $s_{i,\alpha'}=\sum_{\beta=x,y,z}R^{\alpha\beta}s_{i,\beta}$
, as was done in a different context in Ref. \onlinecite{kacsch07epjb}. }

The second term in Eq. (\ref{eq:DDIAssemblyEnergy}) is the dipole-dipole
interaction (DDI) between nanoclusters which can be written as ${\cal E}_{i}^{\mathrm{DDI}}=\xi\sum_{j<i}{\bf s}_{i}\cdot{\cal D}_{ij}\cdot{\bf s}_{j}$,
where ${\cal D}_{ij}$ is the DDI tensor ${\cal D}_{ij}\equiv\frac{1}{r_{ij}^{3}}\left(3{\bf e}_{ij}{\bf e}_{ij}-1\right)$,
with ${\bf r}_{ij}={\bf r}_{i}-{\bf r}_{j}$ and ${\bf e}_{ij}={\bf r}_{ij}/r_{ij}$
is the unit vector along the $i$--$j$ bond.

For convenience, we have introduced the following dimensionless parameters
\begin{eqnarray*}
x & \equiv & \frac{mH}{k_{B}T},\ \sigma\equiv\frac{K_{2}V}{k_{B}T},\ \zeta\equiv\frac{K_{4}}{K_{2}},\ \xi\equiv\left(\frac{\mu_{0}}{4\pi}\right)\left(\frac{m^{2}/a^{3}}{k_{B}T}\right)
\end{eqnarray*}
 together with the DDI coefficient $\tilde{\xi}\equiv\xi\mathcal{C}^{\left(0,0\right)}$.
$\mathcal{C}^{\left(0,0\right)}=-4\pi\left(D_{z}-\frac{1}{3}\right)$
and $D_{z}$ is the demagnetizing factor along the $z$ axis. $K_{2},K_{4}$
are the constants of the uniaxial and cubic anisotropy, respectively.
$a$ is the ``super-lattice'' parameter or the inter-particle distance
in the assembly whose particles are supposed to occupy a simple cubic
(SC) lattice. \textcolor{black}{Yet, we stress that a generalization
to other super-lattices (FCC, BCC, ...) is rather straightforward.
One should simply re-evaluate the lattice sums $\mathcal{C}^{\left(0,0\right)}$
for the given super-lattice. Similarly, one could easily mimic a disordered
spatial arrangement by an evaluation of $\mathcal{C}^{\left(0,0\right)}$
in the case of a randomly depleted lattice. However, for the sake
of clarity and to keep our discussion simple we will consider the
SC case in the rest of this paper.}

The (dimensionless) DDI field $\bm{\Xi}_{i}$ acting on the magnetic
moment $\bm{m}_{i}$ reads 
\begin{equation}
\bm{\Xi}_{i}=\xi\sum_{j}\mathcal{D}_{ij}\cdot\mathbf{s}_{j}.\label{ZetaEffectiveField}
\end{equation}
Later we make use of the spin average $\left\langle \Xi_{i,\parallel}^{2}\right\rangle _{0}$,
where $\Xi_{i,\parallel}=\bm{\Xi}_{i}\cdot\mathbf{e}_{i}$ is the
longitudinal component of $\bm{\Xi}_{i}$, which is defined by\textcolor{black}{
\begin{equation}
{\color{red}{\color{black}\left\langle \Xi_{i,\parallel}^{2}\right\rangle _{0}\equiv\frac{1}{4\pi}\int\left(\prod_{j}d\bm{s}_{j}\right)\Xi_{i,\parallel}^{2}\, e^{\sum_{j}{\cal E}_{j}^{(0)}}.}}\label{eq:SpinAverage}
\end{equation}
}The average $\left\langle {}\right\rangle _{0}$ is defined with
respect to the Gibbs probability distribution containing only the
energy contributions pertaining to a free cluster. Finally, the spin
average of the transverse component of $\bm{\Xi}_{i}$ can be obtained
from the identity $\left\langle \Xi_{i,\perp}^{2}\right\rangle _{0}=\left\langle \Xi_{i}^{2}\right\rangle _{0}-\left\langle \Xi_{i,\parallel}^{2}\right\rangle _{0}$.

\subsection{Statement of the problem}

In the present work we shall be concerned with the study of the combined
effects of surface anisotropy and dipolar interactions on the dynamic
susceptibility of an assembly of monodisperse nanoclusters with oriented
uniaxial anisotropy. The cubic anisotropy which stems from spin non-collinearities
on the cluster's surface is assumed to have its axes parallel to the
crystal axes. We then derive analytical formulas in several cases
of low field ($x\ll1$), high-energy barrier ($\sigma\gg1$), small
surface anisotropy ($\left|\zeta\right|<1$) and weak DDI ($\xi\ll1$).
In particular, for the calculation of the spin averages (\ref{eq:SpinAverage})
and kindred ones we will drop all terms of orders higher than $2$.
For this reason, it turns out that the calculation of such averages
can be done with good approximation with only the uniaxial anisotropy
contribution in the Gibbs probability distribution.\citep{sabsabietal13prb}
The final results are expressed in the end in terms of the following
well known averages (obtained in the absence of a magnetic field)
$\left\langle s_{i}^{\alpha}\right\rangle _{0}=0$, and 

\begin{equation}
\left\langle s_{j}^{\alpha}s_{k}^{\beta}\right\rangle _{0}=\left[\frac{1}{3}(1-S_{j2})\delta^{\alpha\beta}+S_{j2}e_{j}^{\alpha}e_{j}^{\beta}\right]\delta_{jk}\label{firstsecondaverages}
\end{equation}
with\citep{raiste97prb,jongar01prb} 
\begin{equation}
S_{il}(\sigma_{i})\simeq{\displaystyle \left\lbrace \begin{array}{ll}
\frac{(l-1)!!}{(2l+1)!!}(\frac{\sigma_{i}}{2})^{l/2}+\ldots, & \sigma_{i}\ll1,\\
\\
1-\frac{l(l+1)}{4\sigma_{i}}+\ldots, & \sigma_{i}\gg1.
\end{array}\right.}\label{eq:AnisotropyFunction}
\end{equation}

\subsection{Equilibrium susceptibility}

For a weakly interacting assembly of nanoclusters described with the
help of the EOSP model, the equilibrium susceptibility reads (to first
order in $\tilde{\xi}$)

\begin{equation}
\chi^{\mathrm{eq}}\left(x,\sigma,\zeta,\tilde{\xi}\right)\simeq\chi_{\mathrm{free}}^{\mathrm{eq}}+\tilde{\xi}\chi_{\mathrm{int}}^{\mathrm{eq}}\label{eq:XiEq}
\end{equation}
where $\chi_{\mathrm{free}}^{\mathrm{eq}}$ is the equilibrium (linear)
susceptibility of the non-interacting assembly in the limit of high
anisotropy energy barrier\citep{sabsabietal13prb,jongar01prb}
\begin{eqnarray}
\chi_{\mathrm{free}}^{\mathrm{eq}}\left(x,\sigma,\zeta\right) & = & 2\chi_{0}^{\perp}\sigma\left[\chi_{\mathrm{free}}^{\left(1\right)}+3\chi_{\mathrm{free}}^{\left(3\right)}x^{2}\right],\label{eq:XiEqFree}\\
\nonumber \\
\chi_{\mathrm{free}}^{\left(1\right)} & = & \left(1-\frac{1}{\sigma}\right)+\frac{\zeta}{\sigma}\left(-1+\frac{2}{\sigma}\right),\nonumber \\
\chi_{\mathrm{free}}^{\left(3\right)} & = & \frac{1}{3}\left[\left(-1+\frac{2}{\sigma}\right)+\frac{\zeta}{\sigma}\left(2-\frac{5}{\sigma}\right)\right].\nonumber 
\end{eqnarray}
Here $\chi_{0}^{\perp}$ is the transverse equilibrium susceptibility
per spin at zero temperature in the absence of a bias field
\[
\chi_{0}^{\perp}\equiv\left(\frac{\mu_{0}m^{2}}{2K_{2}V}\right).
\]
This can be obtained from Eq. (3.86) of Ref. \onlinecite{garpal00acp}
upon setting the field to zero. 

The contribution of DDI to the equilibrium susceptibility is given
by\citep{sabsabietal13prb}
\begin{eqnarray}
\chi_{\mathrm{int}}^{\mathrm{eq}}\left(x,\sigma,\zeta\right) & = & 2\chi_{0}^{\perp}\sigma\left[\chi_{\mathrm{int}}^{\left(1\right)}+3\chi_{\mathrm{int}}^{\left(3\right)}x^{2}\right],\label{eq:XiEqIntContr}\\
\nonumber \\
\chi_{\mathrm{int}}^{\left(1\right)} & = & 1-\frac{2}{\sigma}-2\left(1-\frac{3}{\sigma}\right)\frac{\zeta}{\sigma},\nonumber \\
\chi_{\mathrm{int}}^{\left(3\right)} & = & -\frac{4}{3}\left[\left(1-\frac{3}{\sigma}\right)-\frac{3\zeta}{\sigma}\right].\nonumber 
\end{eqnarray}
In the sequel, all susceptibilities will be measured in units of $\chi_{0}^{\perp}$.

\section{AC Susceptibility}

The dynamic response of the EOSP assembly can be studied with the
help of the AC susceptibility. For an arbitrary angle $\psi$ between
the (common) easy axis and the field direction, the effective susceptibility
may be written as $\chi=\chi_{\parallel}\cos^{2}\psi+\chi_{\perp}\sin^{2}\psi$. 

Shliomis and Stepanov\citep{shlste93jmmm} proposed a simple Debye
form for $\chi(\omega)$ which can be generalized to describe the
effect of a longitudinal bias field by writing 
\begin{equation}
\chi=\dfrac{\chi_{\parallel}(T,H)}{1+i\omega\tau_{\parallel}}\cos^{2}\psi+\dfrac{\chi_{\perp}(T,H)}{1+i\omega\tau_{\perp}}\sin^{2}\psi,\label{eq:chi_SHS}
\end{equation}
where $\tau_{\parallel}$ and $\tau_{\perp}$ are appropriate longitudinal
(inter-well) and transverse (intra-well) relaxation times; $\chi_{\parallel}(T,H)$
and $\chi_{\perp}(T,H)$ are respectively the longitudinal and transverse
components of the equilibrium susceptibility.

For an assembly with oriented anisotropy in a longitudinal field ($\psi=0$),
we may assume that the transverse response is instantaneous, \emph{i.e.}
$\tau_{\perp}=0$. In this case the AC susceptibility is given by
Eq.~(\ref{eq:chi_SHS}) or using $\tau_{\parallel}=\Gamma^{-1}$
and $\chi_{\parallel}=\chi^{\mathrm{eq}}$ defined in Eq. (\ref{eq:XiEq}),
\begin{equation}
\chi\left(x,\sigma,\zeta,\tilde{\xi},\eta\right)=\frac{\chi^{\mathrm{eq}}}{1+i\omega\Gamma^{-1}}.\label{eq:XiACAssembly}
\end{equation}

Next, we introduce the reduced frequency 
\begin{equation}
\eta\left(x,\sigma,\zeta,\tilde{\xi},\lambda\right)\equiv\omega\tau_{\parallel}=\left(\omega\tau_{D}\right)\left(\tau_{D}\Gamma\right)^{-1},\label{eq:RedFreq}
\end{equation}
with $\lambda$ being the damping parameter. $\Gamma\left(x,\sigma,\zeta,\tilde{\xi},\lambda\right)$
is the relaxation rate of an EOSP nanocluster weakly interacting within
the assembly;\textcolor{red}{{} ${\color{black}\tau_{D}=\left(\lambda\gamma_{{\rm gyr.}}H_{K}\right)^{-1}}$}
is the free diffusion time, $H_{K}=2K_{2}V/M$ the (uniaxial) anisotropy
field, and \textcolor{red}{${\color{black}\gamma_{{\rm gyr.}}\simeq1.76\times10^{11}}$}
(T.s)$^{-1}$ the gyromagnetic ratio. For example, for cobalt particles
the anisotropy field is $H_{K}\sim0.3$ T, and for $\lambda=0.1-10$,
$\tau_{D}\sim2\times10^{-10}-2\times10^{-12}$ s.

At this point, the only missing ingredient to evaluate the susceptibility
in Eq. (\ref{eq:XiACAssembly}) is the relaxation rate. Therefore,
the next Section is devoted to the calculation of the relaxation rate
$\Gamma\left(x,\sigma,\zeta,\tilde{\xi},\lambda\right)$.

\subsection{\label{sub:Relaxation-rate}Relaxation rate}

Here we derive an expression for the relaxation rate of a weakly interacting
EOSP nanocluster.

In Ref.~\onlinecite{jongar01epl} Jönsson and Garcia-Palacios derived
the following approximate expression for $\Gamma$ 
\begin{equation}
\Gamma\simeq\Gamma_{0}\left[1+\dfrac{1}{2}\left\langle \Xi_{\parallel}^{2}\right\rangle _{0}+\dfrac{1}{4}F(\alpha)\left\langle \Xi_{\perp}^{2}\right\rangle _{0}\right].\label{eq:RRDDI}
\end{equation}

This takes account of the various approximations stated earlier inasmuch
as the general spin averages $\left\langle ...\right\rangle $ are
replaced by their analogs $\left\langle ...\right\rangle _{0}$ defined
in Eq. (\ref{eq:SpinAverage}). $\Gamma_{0}$ is the relaxation rate
in the absence of DDI. The function $F(\alpha)$ is given by\citep{garetal99pre}
\begin{equation}
F(\alpha)=1+2(2\alpha^{2}e)^{1/(2\alpha^{2})}\gamma(1+\dfrac{1}{2\alpha^{2}},\dfrac{1}{2\alpha^{2}}),\label{eq:DampingFunction}
\end{equation}
 with $\gamma(a,z)=\int_{0}^{z}dt\, t^{a-1}e^{-t}$ the incomplete
gamma function, and where $\alpha=\lambda\sqrt{\sigma}$. \textcolor{black}{In
}Ref.~\onlinecite{jongar01epl}\textcolor{black}{{} the free-particle
relaxation rate $\Gamma_{0}$ was given in the absence of the applied
magnetic field, }\textcolor{black}{\emph{i.e.}}\textcolor{black}{{}
$\tau_{D}\Gamma_{0}=\frac{2}{\sqrt{\pi}}\sigma^{1/2}e^{-\sigma}$.}\textcolor{red}{{}
}\textcolor{black}{A more general expression for the free-particle
relaxation rate in a longitudinal magnetic field is the Néel-Brown
formula\citep{aha69pr}}

\textcolor{black}{
\begin{equation}
\begin{array}{lll}
\tau_{D}\Gamma_{{\rm NB}} & = & \dfrac{\sigma^{1/2}\left(1-h^{2}\right)}{\sqrt{\pi}}\\
 &  & \times\left[\left(1+h\right)e^{-\sigma\left(1+h\right)^{2}}+\left(1-h\right)e^{-\sigma\left(1-h\right)^{2}}\right],
\end{array}\label{eq:RRFreeAssembly}
\end{equation}
with $h\equiv x/2\sigma$. Setting $h=0$ recovers the previous expression.}

The relaxation rate (\ref{eq:RRFreeAssembly}) has to be generalized
for the present purposes in order to take into account surface anisotropy,
in addition to the magnetic field as well as the core anisotropy. 

For intermediate-to-high damping Langer's approach allows us to compute
the relaxation rate $\Gamma$ of a system with many degrees of freedom
related with its transition from a metastable state through a saddle
point\citep{lan68prl,lan69ap,bra94jap,bra94prb,kac03epl,kac04jml}

\begin{equation}
\Gamma=\frac{\left|\kappa\right|}{2\pi}\frac{\tilde{\mathcal{Z}}_{s}}{\mathcal{Z}_{m}},\label{eq:LangerRR}
\end{equation}
where $\mathcal{Z}_{m}$ and $\tilde{\mathcal{Z}}_{s}$ are respectively
the partition functions in the vicinity of the energy metastable minimum
and the saddle point. The two partition functions are computed using
a quadratic expansion of the energy at the corresponding stationary
states. The attempt frequency $\kappa$ is computed upon linearizing
the dynamical equation around the saddle point, diagonalizing the
resulting matrix and selecting its negative eigenvalue.\citep{lan68prl,lan69ap}

\begin{figure}
\begin{centering}
\includegraphics[width=0.53\columnwidth]{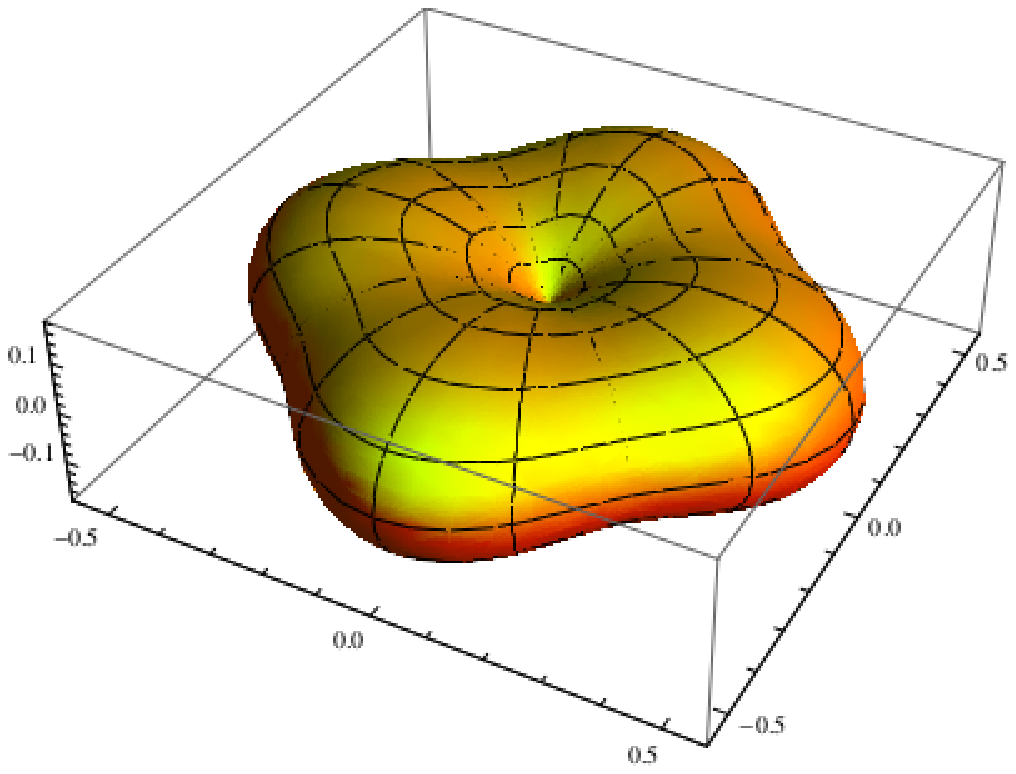}\includegraphics[width=0.45\columnwidth]{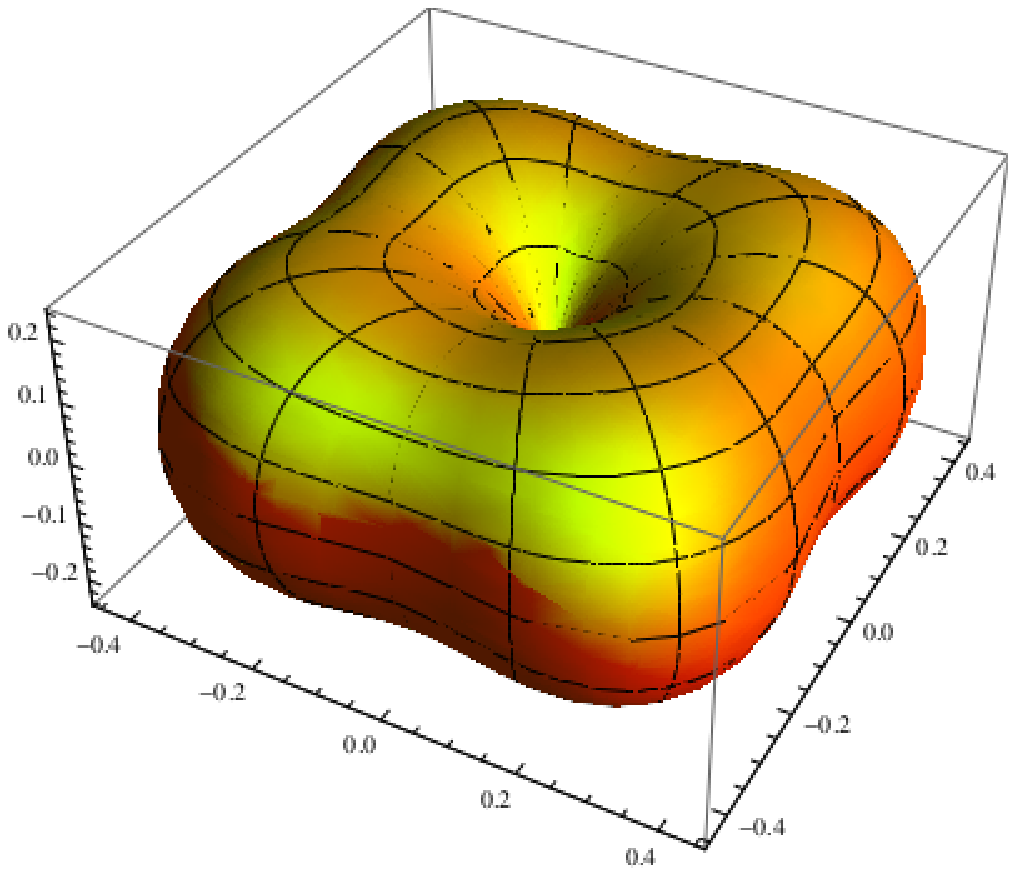}
\par\end{centering}

\caption{\label{fig:SmallZeta}Energy landscape at zero field in the limit
of a large uniaxial anisotropy, for $\zeta>0$ (left), and $\zeta<0$
(right).}
\end{figure}

The dynamics of a single magnetic moment is governed by the (damped)
Landau-Lifshitz equation and Langer's (or Néel-Brown) expression renders
the relaxation rate for its escape from the minimum $(\theta^{(m)},\varphi^{(m)})$
through the saddle point $\left(\theta^{(s)},\varphi^{(s)}\right)$,
in the limit of intermediate-to-high damping. Owing to the approximations
adopted in this work, especially the smallness of the surface anisotropy
with respect to the uniaxial anisotropy ($\left|\zeta\right|<1$),
the energy potential of the non-interacting cluster presents two global
minima that are mainly defined by the uniaxial anisotropy, as is shown
in Fig. \ref{fig:SmallZeta} (in zero field), while the surface anisotropy
induces saddle points at the equator. In the present case, changing
the sign of $\zeta$ does not affect the loci of the minima but those
of the saddle points are rotated by $\pi/4$ around the $z$ axis.
The overall shape of the energy landscape remains, though, quite similar.
The global minima are $\theta^{(m)}=0,\pi$ with uniaxial symmetry
around the $z$ axis. Then, we have
\begin{equation}
\mathcal{Z}_{m}\simeq\frac{2\pi}{2\sigma\left(1-\zeta-h\right)}e^{\mathcal{E}_{m}^{\left(0\right)}},\label{eq:Zm}
\end{equation}
where $\mathcal{E}_{m}^{\left(0\right)}=2\sigma\times\frac{1}{4}\left(2-\zeta-4h\right)$
is the energy at the metastable minimum $\theta^{(m)}=\pi$. 

There are four equivalent escape routes (saddle points) related to
each other by a rotational symmetry with respect to the azimuthal
angle $\varphi$ 
 and their loci depend on the sign of $\zeta$. Indeed, for $\zeta>0$
we have $\varphi^{(s)}=\frac{\pi}{4},\frac{3\pi}{4},\frac{5\pi}{4},\frac{7\pi}{4}$
and

\begin{equation}
\cos\theta^{(s)}=\sqrt{\frac{2+\zeta}{3\zeta}}\sin\left(\frac{\phi}{3}\right)-\sqrt{\frac{2+\zeta}{9\zeta}}\cos\left(\frac{\phi}{3}\right)\label{eq:SPz>0theta}
\end{equation}
with $\cos\phi=\frac{9h\zeta^{1/2}}{\left(2+\zeta\right)^{3/2}}$.

For $\zeta<0$ the saddle points are given by $\varphi^{(s)}=0,\frac{\pi}{2},\pi,\frac{3\pi}{2}$
and 
\begin{equation}
\cos\theta^{(s)}=\left[\frac{h}{4\zeta}+\sqrt{\Delta}\right]^{1/3}+\left[\frac{h}{4\zeta}-\sqrt{\Delta}\right]^{1/3}\label{eq:SPz<0theta}
\end{equation}
with $\Delta=\left(\frac{h}{4\zeta}\right)^{2}-\left(\frac{1+\zeta}{6\zeta}\right)^{3}$.

\begin{figure}
\begin{centering}
\includegraphics[width=0.95\columnwidth]{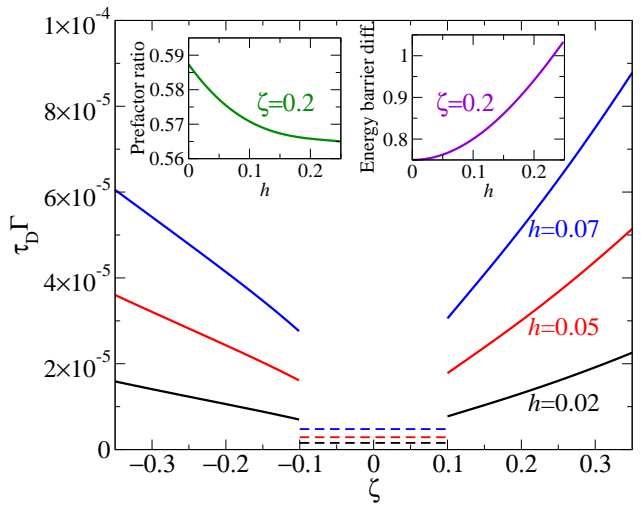}
\par\end{centering}

\caption{\label{fig:RR}Relaxation rate as a function of the (negative and
positive) parameter $\zeta$, for three values of the (reduced) applied
field $h$: \textcolor{black}{Full lines correspond to the relaxation
rate defined in Eq. (\ref{eq:prefactor_def}), while the dashed lines
are plots of the $\zeta-$independent Néel-Brown relaxation rate (\ref{eq:RRFreeAssembly})
for different values of the magnetic field. Insets: Ratio of the two
prefactors $\Gamma_{p}^{+}\left(h,\sigma,\zeta,\lambda\right)/\Gamma_{p}^{-}\left(h,\sigma,\zeta,\lambda\right)$,
as defined in Eq. (\ref{eq:prefactor_def}), and energy barrier difference
against $h$, for $\sigma=15$ and $\zeta=0.2$.}}
\end{figure}

\textcolor{black}{For a small magnetic field $h$, the azimuthal angle
at the saddle point remains close to the equator while an expansion
of Eq. (\ref{eq:SPz>0theta}) yields $\theta^{\left(s\right)}\simeq\frac{\pi}{2}+\frac{2h}{2+\zeta}$.
It is worth mentioning that the symmetry breaking of the continuous
rotation around $\varphi$, induced by the introduction of a cubic
anisotropy, appears as soon as $\zeta$ assumes a finite value. However,
for very small values of $\zeta$ the energy surface around the saddle
points remains flat rendering the quadratic expansion of the energy
at the saddle point questionable {[}see Fig.}\textcolor{red}{{} }\textcolor{black}{\ref{fig:RR}
below{]}. As a consequence Langer's approach does not apply in such
situations, as was emphasized earlier.\citep{titovetal05prb,dejardinetal08jpd} }

Next, expanding the energy at the saddle points for $\zeta>0$ and
$\zeta<0$ (with not too small $\left|\zeta\right|$) we obtain the
following generic expression for the relaxation rate (upon multiplying
by the symmetry factor $4$) $\Gamma_{0}=\Gamma_{\left(\pi,0\right)\rightarrow\left(\theta^{(s)},\varphi^{(s)}\right)}$
\begin{equation}
\tau_{D}\Gamma_{0}=4\times\frac{\left|\kappa\right|}{2\pi}\sin\theta^{(s)}\frac{2\sigma\left(1-\zeta-h\right)}{\sqrt{\left|\mu_{1}^{\left(s\right)}\mu_{2}^{\left(s\right)}\right|}}e^{\Delta\mathcal{E}^{\left(0\right)}}.\label{eq:GammaZeta}
\end{equation}
The attempt frequency $\kappa$, as a function of the damping parameter
$\lambda$, is given by the general expression
\begin{equation}
\begin{array}{lll}
\kappa & = & \frac{\lambda}{2}\times\left[\left(\mu_{2}^{(s)}+\mu_{1}^{(s)}\right)\right.\\
 &  & \left.-\sqrt{\left(\mu_{2}^{(s)}+\mu_{1}^{(s)}\right)^{2}-4\left(1+\frac{1}{\lambda^{2}}\right)\mu_{1}^{(s)}\mu_{2}^{(s)}}\right]
\end{array}\label{eq:KappaGeneral}
\end{equation}
where $\mu_{i}^{\left(s\right)},i=1,2$ are the eigenvalues of the
energy quadratic form near the saddle point, with respect to the variables
$\theta,\varphi$, respectively. These, together with the energy at
the saddle point, are given by 
\[
\begin{array}{l}
\mathcal{E}_{s}^{\left(0\right)}=2\sigma\left[h\cos\theta^{(s)}+\frac{1}{2}\cos^{2}\theta^{(s)}-\frac{\zeta}{8}\left(\sin^{4}\theta^{(s)}+2\cos^{4}\theta^{(s)}\right)\right],\\
\\
\mu_{1}^{\left(s\right)}=2\sigma\times\frac{-1}{4}\left[4h\cos\theta^{(s)}+\left(4-\zeta\right)\cos2\theta^{(s)}-3\zeta\cos4\theta^{(s)}\right],\\
\\
\mu_{2}^{\left(s\right)}=2\sigma\left[-\zeta\sin^{4}\theta^{(s)}\right].
\end{array}
\]
for $\zeta>0$. 

As the energy landscape remains globally the same by changing $\zeta\to-\zeta$,
only the energy at the saddle points and the eigenvalues change, yet
the overall form of the relaxation rate is still given by Eqs. (\ref{eq:GammaZeta})
and (\ref{eq:KappaGeneral}) with the following substitutions 
\[
\begin{array}{l}
\mathcal{E}_{s}^{\left(0\right)}=2\sigma\left[h\cos\theta^{(s)}+\frac{1}{2}\cos^{2}\theta^{(s)}-\frac{\zeta}{4}\left(\cos^{4}\theta^{(s)}+\sin^{4}\theta^{(s)}\right)\right],\\
\\
\mu_{1}^{\left(s\right)}=2\sigma\left[-h\cos\theta^{(s)}-\cos\left(2\theta^{(s)}\right)+\zeta\cos\left(4\theta^{(s)}\right)\right],\\
\\
\mu_{2}^{\left(s\right)}=2\sigma\left[\zeta\sin^{4}\theta^{(s)}\right].
\end{array}
\]
for $\zeta<0$.

Finally, the energy barrier $\Delta\mathcal{E}^{\left(0\right)}$
in Eq. (\ref{eq:GammaZeta}) is defined as $\Delta\mathcal{E}^{\left(0\right)}=\mathcal{E}_{s}^{\left(0\right)}-\mathcal{E}_{m}^{\left(0\right)}$.

In the limit of zero field ($h=0$) and for $\zeta>0$, for instance,
$\mu_{1}^{(s)}/2\sigma\rightarrow\left(\zeta+2\right)/2,\,\mu_{2}^{\left(s\right)}/2\sigma\rightarrow\zeta,\,\mathcal{E}_{0}^{(s)}/2\sigma\rightarrow-\zeta/8$
so that the relaxation rate in (\ref{eq:GammaZeta}) reduces to the
result obtained in Ref. \onlinecite{dejardinetal08jpd}, normalized
with respect to the Néel's free-diffusion relaxation time\citep{garlaz98prb}
$\tau_{\mathrm{N}}=\frac{m}{2\alpha\gamma k_{\mathrm{B}}T}=\sigma\tau_{{\rm D}}$. 

Two remarks are in order: 
\begin{itemize}
\item There are two limits to the range of $\zeta$ ($>0$). First, $\zeta$
must not exceed some value that marks the limit of validity of the
EOSP model. From numerical calculations,\citep{kacbon06prb,yanesetal07prb}
this has been evaluated to $\sim0.25$ for an SC lattice and $\sim0.35$
for an FCC lattice. The second limit stems from the fact that the
analytical expressions obtained above for $\Gamma$ within Langer's
approach cannot be continued to $\zeta=0$ because the saddle points
created by the cubic contribution to the anisotropy disappear at the
uniaxial anisotropy limit.\textcolor{red}{{} }\textcolor{black}{The
lower limit on $\zeta$ can be obtained by setting to zero the first
derivative of $\Gamma$ with respect to $\zeta$ and numerically solving
the ensuing equation. Doing so, we find that for $\sigma=15...25,$
for instance, $\zeta_{{\rm crit}}$ is of the order of 0.1.}
\item Because of the non-axial symmetry (owing to the presence of surface
cubic anisotropy) considered here, the relaxation rate depends in
a non trivial way on the damping parameter. Consequently, the longitudinal
response (in-phase and out-of-phase) are damping-dependent.
\end{itemize}
In Fig. \ref{fig:RR} we plot the relaxation rate for both $\zeta>0$
and $\zeta<0$ as a function of $\zeta$ and different values of the
applied field $h$,\textcolor{black}{{} for $\sigma=15$. In this case,
as mentioned above, the relaxation rate computed within our approach
is only valid for $0.1<\left|\zeta\right|<1$. For smaller values
of $\left|\zeta\right|$ Langer's approach is no longer valid and
the relaxation rate is given by the Néel-Brown formula (\ref{eq:RRFreeAssembly})
which does not depend on $\zeta$. This is shown by the dashed lines
in Fig. \ref{fig:RR}. As it can be expected, the relaxation rate
that includes the cubic anisotropy is larger than the Néel-Brown relaxation
rate since the creation of saddle points increases the probability
of escaping from the metastable state.}

Next, if we write the relaxation rates given by Eq. (\ref{eq:GammaZeta})
in the form
\begin{equation}
\Gamma_{0}\left(h,\sigma,\zeta,\lambda\right)=\Gamma_{p}^{\epsilon}\left(h,\sigma,\zeta,\lambda\right)e^{\Delta\mathcal{E}_{\epsilon}^{\left(0\right)}\left(\zeta\right)}\label{eq:prefactor_def}
\end{equation}
with $\epsilon=+$ for $\zeta>0$ and $\epsilon=-$ for $\zeta<0$
we can study the behavior of the ratio of the prefactors and the difference
of the energy barriers as the field is varied. The corresponding plots
are given in the inset in Fig. \ref{fig:RR}. We see that the ratio
of the prefactors is a decreasing function of $h$ while the difference
of the energy barriers is an increasing function thereof. This implies
that there is a competition between the prefactor-dominated dynamics
and the relaxation through the energy-barrier crossing, or in other
words, between the dynamics dominated respectively by the fluctuations
of the transverse and the longitudinal components of the magnetic
moment.

Caution is necessary when trying to compare the expression of the
relaxation rate $\Gamma_{0}\left(h,\sigma,\zeta,\lambda\right)$ derived
here in the presence of both surface effects $\left(\zeta\ne0\right)$
and DDI $\left(\xi\ne0\right)$ with the relaxation rate obtained,
in the absence of the cubic anisotropy, by other authors.\citep{jongar01epl,dejardin_pmd_JAP_2011}\textcolor{blue}{{}
}\textcolor{black}{Indeed, in the presence of an arbitrary magnetic
field, one cannot simply set $\zeta=0$ in our expressions because
these have been derived using Langer's approach that relies on the
validity of the quadratic expansion of the energy at the minima and
saddle points ; a validity that breaks down for rather small (but
non vanishing) values of $\zeta$.}\textcolor{red}{{} }\textcolor{black}{From
a mathematical point of view, setting $\zeta$ to zero in Eqs. (\ref{eq:SPz>0theta},
\ref{eq:SPz<0theta}), for example, leads to a singularity.}

Now, for the assembly we use the spin averages $\left\langle \Xi_{i,\parallel}^{2}\right\rangle _{0}$and
$\left\langle \Xi_{i,\perp}^{2}\right\rangle _{0}$ obtained in Ref.
\onlinecite{jongar01epl} for a monodisperse assembly on a SC lattice
and in the absence of an external magnetic field 
\begin{eqnarray}
\left\langle \Xi_{i,\parallel}^{2}\right\rangle _{0} & = & \frac{\xi^{2}}{3}\left[\left(1-S_{2}\right)\mathcal{R}+3S_{2}\mathcal{T}\right],\nonumber \\
\left\langle \Xi_{i,\perp}^{2}\right\rangle _{0} & = & \frac{\xi^{2}}{3}\left[\left(2+S_{2}\right)\mathcal{R}-3S_{2}\mathcal{T}\right],\label{eq:DDIFieldSquared}
\end{eqnarray}
where $S_{2}$ is defined in Eq. (\ref{eq:AnisotropyFunction}). $\mathcal{R}$
and $\mathcal{T}$ are lattice sums given by $\mathcal{R}=2\sum_{j\neq i}r_{ij}^{-6}$,
$\mathcal{T}=\sum_{j\neq i}\left(\mathbf{e}\cdot{\cal D}_{ij}\mathbf{e}\right)^{2}$.
For a simple cubic lattice we have, in the thermodynamic limit, $\mathcal{R}\simeq16.8,\mathcal{T}\simeq13.4$.

Therefore, using Eqs. (\ref{eq:DampingFunction}), (\ref{eq:GammaZeta}),
and (\ref{eq:DDIFieldSquared}) in Eq. (\ref{eq:RRDDI}) we obtain
the relaxation rate for an assembly of interacting clusters within
the EOSP approach

\begin{equation}
\Gamma\left(h,\sigma,\zeta,\lambda,\xi\right)\simeq\Gamma_{0}\left(h,\sigma,\zeta,\lambda\right)\left[1+\frac{\xi^{2}}{6}\mathcal{S}\left(\lambda\right)\right].\label{eq:RRDDI-EOSP}
\end{equation}
where $\mathcal{S}\left(\lambda\right)$ is defined by
\begin{equation}
\mathcal{S}\left(\lambda\right)=\left(1+F\left(\lambda\right)\right)\mathcal{R}+\left(3\mathcal{T}-\mathcal{R}\right)\left(1-\dfrac{F\left(\lambda\right)}{2}\right)S_{2}.\label{eq:F_of_lambda}
\end{equation}
Alternatively, using $\eta_{0}=\omega\Gamma_{0}^{-1}$, we can also
rewrite Eq. (\ref{eq:RedFreq}) as 
\begin{equation}
\eta\left(h,\sigma,\zeta,\xi,\lambda\right)=\eta_{0}\left(h,\sigma,\zeta,\lambda\right)\left[1+\frac{\xi^{2}}{6}\mathcal{S}\left(\lambda\right)\right].\label{eq:RRDDI-EOSP-freq}
\end{equation}

\subsection{AC susceptibility}

We rewrite the AC susceptibility (\ref{eq:XiACAssembly}) separating
its real and imaginary parts $\chi\left(h,\sigma,\zeta,\tilde{\xi},\eta\right)=\chi^{\prime}-i\chi^{\prime\prime}$
with 
\begin{eqnarray}
\chi^{\prime} & = & \chi^{\mathrm{eq}}\frac{1}{1+\eta^{2}},\qquad\chi^{\prime\prime}=\chi^{\mathrm{eq}}\frac{\eta}{1+\eta^{2}}.\label{eq:ReImXiAssembly}
\end{eqnarray}
Now, we substitute for $\chi^{\mathrm{eq}}$ and $\eta$ their respective
expressions (\ref{eq:XiEq}) and (\ref{eq:RRDDI-EOSP-freq}), taking
account of DDI and surface anisotropy contributions given above. We
obtain

\begin{eqnarray*}
\chi^{\prime} & \simeq & \chi_{\mathrm{free}}^{\prime}+\frac{\xi}{1+\eta_{0}^{2}}\left[\Lambda^{\left(1\right)}+\xi\frac{\eta_{0}^{2}}{1+\eta_{0}^{2}}\Lambda^{\left(2\right)}\right],\\
\\
\chi^{\prime\prime} & = & \chi_{\mathrm{free}}^{\prime\prime}+\frac{\xi\eta_{0}}{1+\eta_{0}^{2}}\left[\Lambda^{\left(1\right)}+\xi\frac{1-\eta_{0}^{2}}{1+\eta_{0}^{2}}\Lambda^{\left(2\right)}\right],
\end{eqnarray*}
where we have defined the in-phase and out-of-phase susceptibilities
in the absence of DDI
\[
\chi_{\mathrm{free}}^{\prime}\left(h,\sigma,\zeta,\lambda\right)\equiv\frac{\chi_{\mathrm{free}}^{\mathrm{eq}}}{1+\eta_{0}^{2}},\ \chi_{\mathrm{free}}^{\prime\prime}\left(h,\sigma,\zeta,\lambda\right)\equiv\frac{\eta_{0}\chi_{\mathrm{free}}^{\mathrm{eq}}}{1+\eta_{0}^{2}}
\]
together with the $1^{\mathrm{st}}$- and $2^{\mathrm{nd}}$-order
DDI contributions
\[
\begin{array}{lll}
\Lambda^{\left(1\right)} & \equiv & \chi_{\mathrm{int}}^{\mathrm{eq}}\mathcal{C}^{\left(0,0\right)},\\
\\
\Lambda^{\left(2\right)} & \equiv & \frac{\chi_{\mathrm{free}}^{\mathrm{eq}}}{3}\mathcal{S}\left(\lambda\right).
\end{array}
\]
$\chi_{\mathrm{free}}^{\mathrm{eq}}$ and $\chi_{\mathrm{int}}^{\mathrm{eq}}$
are given by Eqs. (\ref{eq:XiEqFree}, \ref{eq:XiEqIntContr}) and
$\eta_{0}=\omega\Gamma_{0}^{-1}$ by Eq. (\ref{eq:GammaZeta}).

\section{Results}

\subsection{Noninteracting assembly of OSP nanomagnets}

\begin{figure*}
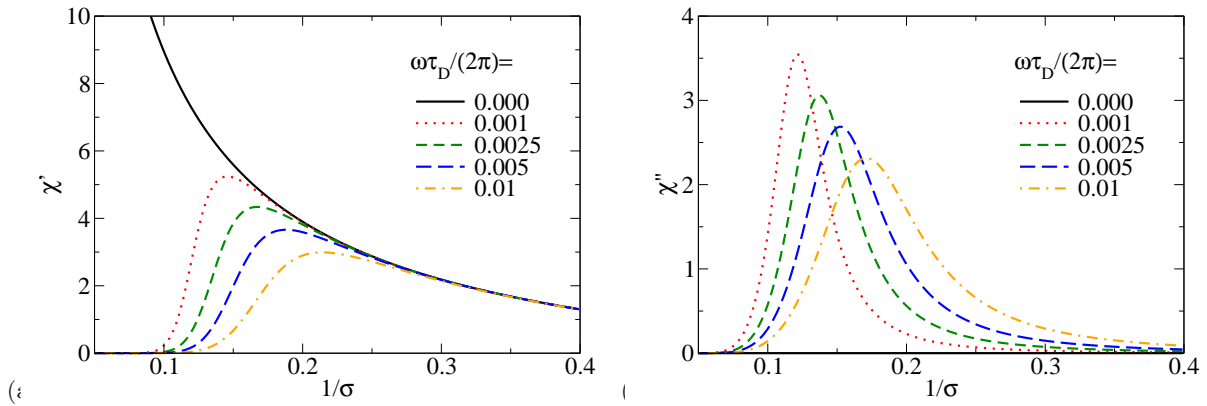

\begin{centering}
(a)\includegraphics[height=0.6\columnwidth]{Fig3a}\quad(b)\includegraphics[height=0.6\columnwidth]{Fig3b}
\par\end{centering}

\caption{\label{fig:ReXiImXi-x=00003D0_xi=00003D0_nuVaried}(a) $\chi^{\prime}$
(left) and (b) $\chi^{\prime\prime}$ for a free assembly ($\xi=0$)
within the OSP model, \emph{i.e.} without surface anisotropy ($\zeta=0$),
for different frequencies $\tilde{f}\equiv\omega\tau_{D}/(2\pi)$.}
\end{figure*}

Using our formalism we first reproduce the well known results for
the in-phase and out-of-phase susceptibilities for an assembly of
noninteracting nanomagnets with uniaxial anisotropy, in zero DC field.\citep{garlaz98prb,garpal00acp}
In Fig. \ref{fig:ReXiImXi-x=00003D0_xi=00003D0_nuVaried} we plot
the in-phase (left) and out-of-phase (right) susceptibilities as functions
of $1/\sigma\propto T$ for zero field ($x=0$) and different frequencies.
On the left we have also included the equilibrium susceptibility $\chi_{\mathrm{free}}^{\mathrm{eq}}\left(\zeta=0\right)$,
represented by the solid line.

The appearance in $\chi^{\prime}$ and $\chi^{\prime\prime}$ of a
maximum at some particular temperature $T_{\max}$ and the displacement
of the latter to the right (higher temperatures) upon increasing the
measuring frequency is already well understood and explained in details,
\emph{e.g.} in Ref. \onlinecite{garlaz98prb}. In particular, the
maximum of $\chi^{\prime}$ is formed as a result of the competition
between the blocking effect (namely the\emph{ }decrease of the relaxation
rate) and the increase of $\chi^{\mathrm{eq}}$ as the temperature
decreases. At low temperature, the relaxation time is longer than
the measuring time $t_{m}=2\pi/\omega$ and thereby over a large number
of cycles of the AC field, the over-barrier switching probability
is nearly zero and the response consists mainly of intra-well rotations.
As $T$ increases the clusters magnetic moments start to depart from
their respective energy minima due to thermal fluctuations. Then,
over the same number of cycles of the AC field the switching probability
acquires a non negligible value. The response starts to increase with
increasing temperature within a range where the thermally-activated
mechanism of over-barrier crossing is not yet efficient enough, leading
to a considerable delay of the response with respect to the excitation.
This leads to a considerable out-of-phase response $\chi^{\prime\prime}$
as witnessed by the increase of the latter, see Fig. \ref{fig:ReXiImXi-x=00003D0_xi=00003D0_nuVaried}
(right). At higher temperatures, the over-barrier crossing mechanism
becomes so efficient that the magnetic moments instantaneously distribute
themselves among the various energy minima, in phase with the probing
field. At much higher temperatures, the distribution of the magnetic
moments reaches its equilibrium state and the $\chi^{\prime}$ curves
become independent of the measuring frequency and superimpose on the
equilibrium linear susceptibility $\chi^{\mathrm{eq}}$, and correspondingly
$\chi^{\prime\prime}$ tends to zero.

The displacement of $T_{\max}$ is easily understood from the expression
of the latter as a function of the measuring frequency $\nu_{m}$.
Indeed, this temperature is related with the over-barrier rotation
process whose relaxation time is approximately given by the simple
Arrhenius law $\tau_{\parallel}=\tau_{0}\exp\left(\Delta E/k_{\mathrm{B}}T\right)$,
where $\Delta E$ is the effective energy barrier and $\tau_{0}\sim10^{-12}-10^{-9}$
s the characteristic time of the intra-well dynamics. At $T=T_{\max}$
we can write $\tau_{\parallel}\simeq t_{\mathrm{m}}$, \emph{i.e.}
the measuring time ($\sim100$ s for a commercial SQUID), and this
then leads to
\begin{equation}
T_{\max}=\frac{\Delta E}{k_{\mathrm{B}}}\times\ln^{-1}\left(\frac{\tau_{{\rm m}}}{\tau_{0}}\right).\label{eq:Tmax}
\end{equation}

From this relation, one can easily infer the increase of $T_{\max}$
as the measuring frequency $\nu_{{\rm m}}=\tau_{{\rm m}}^{-1}$ increases.
From the physical viewpoint, with higher $\nu_{{\rm m}}$ one probes
on average more probable (with higher relaxation rate) switching processes
and this is in effect induced by an increase in temperature.

\subsection{Noninteracting assembly: effects of surface anisotropy}

Now, to investigate the effect of surface anisotropy on the AC susceptibility
we can compute the real and imaginary components of the latter as
functions of temperature, for different values of the parameter $\zeta>0$.

We have observed that the maxima of both $\chi^{\prime}$ and $\chi^{\prime\prime}$
shifts toward higher temperatures as $\zeta$ increases. Indeed, setting
to zero the first derivative of $\chi^{\prime}$ with respect to temperature
and setting $T=T_{\max}$ in the ensuing equation, we can solve the
latter for $T_{\max}$ as a function of the other parameters, especially
$\zeta$. We indeed find a monotonously increasing function of $\zeta$.
Intuitively this result appears to be at variance with the fact that
since the cubic (surface) anisotropy creates saddle points it leads
to an increase of the relaxation rate and thereby to a decrease of
$T_{\max}$. However, as mentioned earlier the location of the maximum
of the dynamic response, while it does depend on the energy barriers,
it is strongly dependent on the equilibrium response (\emph{i.e.}
$\chi^{\mathrm{eq}}$) which is rather different for the pure uniaxial
case ($\zeta=0$). More precisely, $\chi^{\mathrm{eq}}$ is a decreasing
function of $\zeta$ and thereby when $\zeta$ increases the dynamic
response requires higher temperatures to reach its maximum, thus leading
to an increasing $T_{\max}$ for increasing $\zeta$.

\subsection{Effects of inter-particle interactions in the absence of surface
anisotropy}

\begin{figure*}
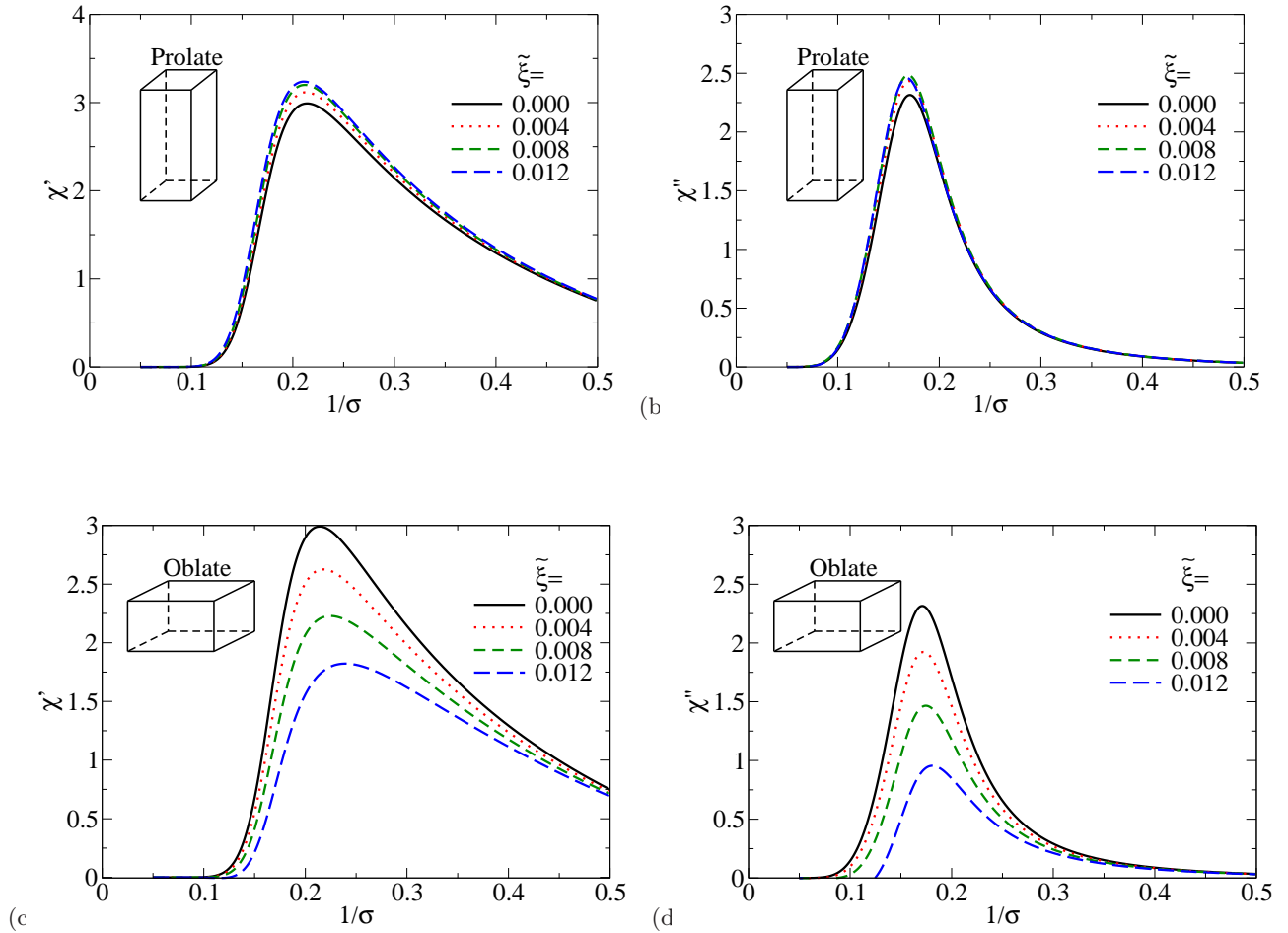

\begin{centering}
(a)\includegraphics[height=5.5cm]{Fig4a}\quad(b)\includegraphics[height=5.5cm]{Fig4b}
\par\end{centering}
$$\ $$
\begin{centering}
(c)\includegraphics[height=5.5cm]{Fig4c}\quad(d)\includegraphics[height=5.5cm]{Fig4d}
\par\end{centering}

\caption{\label{fig:ReImXiVSt_OSP-DDI0.0-0.012}(a) $\chi^{\prime}$ and (b)
$\chi^{\prime\prime}$ for an interacting prolate $\left(10\times10\times20\right)$
assembly with varying DDI strength $\tilde{\xi}$, for the frequency
$\tilde{f}\equiv\omega\tau_{D}/(2\pi)=0.01$ in the absence of an
external field $h=0$, in the high damping regime $\lambda=10$. Same
plots in (c) and (d) for an oblate $\left(20\times20\times5\right)$
assembly.}
\end{figure*}
The effect of DDI on the AC susceptibility has been widely investigated
by many groups.\citep{doretal96prb,anderssonetal97j3m,spinu98phd,mamnak98jmmm,bergor01jpcm,garpalgar04prb,jongar01epl,shimetal06jap,singhetal06jap,singhetal08jap,godselletal08jpd,masunagaetal09prb,ledueetal12jnn}
In Ref. \onlinecite{ledueetal12jnn} the authors provide a short review
of the situation regarding the effect of DDI on the maximum of $\chi^{\prime}$
and $\chi^{\prime\prime}$ and their shift in temperature as the DDI
intensity is varied and the assembly shape changed from oblate to
prolate. It was argued that the discrepancy of conclusions found in
the literature as to whether the DDI shift the maximum of $\chi^{\prime}$
and $\chi^{\prime\prime}$ towards higher or lower temperatures resides
in many reasons, mostly related with the effects of damping, the shape
of the (assembly) sample, and anisotropy. Here we use the same formalism
and approximations and obviously confirm the same results. Therefore,
we shall not repeat the conclusions of the previous work. 

Nevertheless, Fig. \ref{fig:ReImXiVSt_OSP-DDI0.0-0.012} shows that
as the shape of the assembly changes from prolate to oblate, we obtain
an opposite shift in temperature in both the maximum of $\chi^{\prime}$
and $\chi^{\prime\prime}$ and also in the corresponding $T_{\max}$.\textcolor{red}{{}
}\textcolor{black}{In the case of isotropic samples, such as cubes,
the lattice sum $\mathcal{C}^{\left(0,0\right)}$ vanishes leading
to a DDI coefficient $\tilde{\xi}=0$. Therefore, no shift is observed
and the DDI do not contribute, within the present approach. For prolate
and oblate samples, both} shifts are explained by the fact that the
equilibrium susceptibility increases with DDI in a prolate sample
whereas it decreases in an oblate sample. More importantly, it is
seen that the effect of DDI is more pronounced in the oblate case
because there the DDI are in competition with the uniaxial anisotropy
and thus strongly contribute to suppress the equilibrium susceptibility.
The effect of damping, while remaining secondary as compared to that
of the assembly shape, seems to be somewhat more pronounced in the
case of prolate samples. This may be due again to the fact that in
the prolate case the increase of $\chi^{\mathrm{eq}}$ with DDI is
slower than its decrease for the oblate shape. As such, $\chi^{\prime}$
and $\chi^{\prime\prime}$, and more so for $\chi^{\prime}$, are
more sensitive to the change of the relaxation rate which then starts
to prevail, and which does depend on damping.

\subsection{DDI versus surface effects}

Now we are ready to investigate the interplay between inter-particle
DDI and intrinsic surface anisotropy. We only present the case of
$\zeta>0$ in which surface (cubic) anisotropy favors the magnetic
alignment along the cube diagonals. In order to deal with the case
$\zeta<0$ one has to use the corresponding relaxation rate, as discussed
in Section \ref{sub:Relaxation-rate}. Yet, as shown in Fig. \ref{fig:RR}
the behavior of the relaxation rate for $\zeta<0$ is qualitatively
the same as that for $\zeta>0$ and that even quantitatively the difference
is not really significant. Therefore, in the remaining part of the
paper we will focus our discussion on $\zeta>0$. 

We have systematically analyzed $\chi^{\prime}$ and $\chi^{\prime\prime}$
for various values of the surface anisotropy coefficient $\zeta$,
for both prolate and oblate assemblies. We have observed the upward
shift of $T_{\max}$ as $\zeta$ increases and the downward shift
of the maximum of $\chi^{\prime}$ and $\chi^{\prime\prime}$, as
already discussed earlier. However, owing to the fact that the effect
of increasing $\zeta$ is to draw the particle's magnetic moment towards
the cube diagonals, it basically plays the same role in a prolate
sample where the magnetization is enhanced along the $z$ axis, or
in an oblate sample where the magnetization is enhanced in the $xy$
plane.

\begin{figure}
\begin{centering}
\includegraphics[width=0.95\columnwidth]{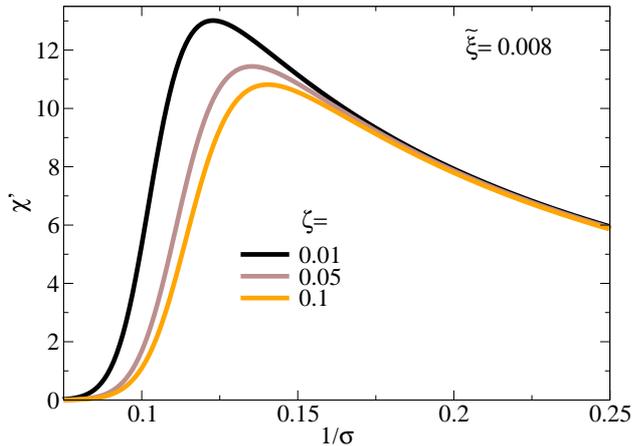}
\par\end{centering}

\caption{\label{fig:compar_DDI_vs_zeta}$\chi^{\prime}$ for an interacting
prolate $\left(10\times10\times20\right)$ assembly with a fixed DDI
strength $\tilde{\xi}=0.008$ and varying surface anisotropy coefficient
$\zeta$, for the frequency $\tilde{f}\equiv\omega\tau_{D}/(2\pi)=0.01$.
$h=0$.}
\end{figure}

The effect of increasing the strength of DDI alone is shown in Fig.
\ref{fig:ReImXiVSt_OSP-DDI0.0-0.012}. In the case of a prolate sample,
we have observed a shift of the maximum toward lower temperatures,
in the absence of surface anisotropy. Note again that this is not
the $\zeta=0$ limit of the expressions of Section \ref{sub:Relaxation-rate}.
It is simply the OSP model with the relaxation rate (\ref{eq:RRFreeAssembly}).
The effect of frequency observed by Lee \emph{et al}. \citep{leeetal11apl}
is similar to the behavior that we observe here: $T_{{\rm max}}$
increases and $\chi^{\prime}$ decreases. Furthermore, the fact that
$T_{{\rm max}}$ increases as the concentration increases is in line
with what we observe for oblate samples and corresponds to the type
of samples investigated by Lee \emph{et al}. Despite the relative
success of our model in interpreting the experimental data, one has
to be careful as not to push the comparison too far because our approach
has been derived for textured monodisperse assemblies and, more importantly,
is perturbative and thus inherently restricted to weak DDI. This is
in general not the case in experiments where the assemblies are often
random and rather dense. In such cases (especially high densities),
a more quantitative comparison with experiments can only be accessible
with the help of numerical investigations\citep{russieretal12jap,brinisetal_JAP2014}.

In Fig. \ref{fig:compar_DDI_vs_zeta} we present a specific case in
order to highlight the competing effects of surface and dipolar interaction
on the susceptibility. The curves are obtained for $\tilde{\xi}=0.008$
and small (and increasing) surface anisotropy parameter $\zeta$.
These results show that the surface anisotropy, in the present case
of positive $\zeta$, has the opposite effect to that of DDI. More
precisely, this implies that surface effects can screen out the effect
of DDI and the other way round. This confirms the results of Ref.
\onlinecite{sabsabietal13prb} for equilibrium properties for both
negative and positive $\zeta$.

\subsection{Discussion}

Very often the experimental results related with the dynamics of an
assembly of DDI-coupled nanoparticles are analyzed with the help of
the Vogel-Fulcher law \citep{masunagaetal09prb,dejardin_pmd_JAP_2011,leeetal11apl,landi13jap,fiorani14jpcs,allia14jpcs}
\begin{equation}
\Gamma=\tau_{0}^{-1}\, e^{\frac{\Delta E}{k_{B}\left(T-\theta_{\mathrm{VF}}\right)}}\label{eq:VF}
\end{equation}
where $\nu_{0}=\tau_{0}^{-1}\simeq10^{9}-10^{12}$ Hz, $\theta_{\mathrm{VF}}$
represents an effective temperature supposed to include the DDI correction
and $\Delta E$ is the energy barrier, which reads $\Delta E=K_{2}V$
in the case of uniaxial anisotropy and zero field. The main concern
with this phenomenological formula is to provide an interpretation
of the parameter $\theta_{\mathrm{VF}}$ on physical grounds. Accordingly,
in Ref. \onlinecite{landi13jap}. there is a discussion of a few approaches
in this regard\emph{. }For instance, it is shown how the work of Shtrikman
and Wohlfarth\citep{shtwoh81pla} leads to an expression of $\theta_{\mathrm{VF}}$
in terms of the applied magnetic field and how the work by Déjardin\citep{dejardin_pmd_JAP_2011}
yields an expression in terms of the DDI coupling. In the work of
Landi itself $\theta_{\mathrm{VF}}$ is expressed in terms of the
inter-particle distance and other parameters such as the particles
magnetic moment and the uniaxial-anisotropy energy. 

Here we show that our formalism is in full agreement with the previous
results and further extends them along the following lines: i) surface
anisotropy, ii) particles spatial distribution and shape of the assembly,
iii) damping parameter.

In Eq. (\ref{eq:RRDDI-EOSP}) the factor $\Gamma_{0}\left(h,\sigma,\zeta,\lambda\right)$
depends on the applied field, surface anisotropy and damping, together
with other parameters, as is seen in Eq. (\ref{eq:prefactor_def}).
It turns out that in fact the prefactor $\Gamma_{p}^{+}\left(h,\sigma,\zeta,\lambda\right)$
is a slowly varying function of $\zeta$ and as such can be written
as $\Gamma_{p}^{+}\left(h,\sigma,\zeta,\lambda\right)\simeq\tilde{\Gamma}\left(h,\sigma,\lambda\right)$.
This implies that the dependence of the relaxation rate $\Gamma_{0}\left(h,\sigma,\zeta,\lambda\right)$
$\zeta$ is mainly borne by the energy barrier $\Delta\mathcal{E}_{+}^{\left(0\right)}\left(\zeta\right)$.
Therefore, in zero field $\Delta\mathcal{E}_{+}^{\left(0\right)}\left(\zeta\right)\simeq-\sigma+\sigma\zeta/4$
and upon expanding in $\zeta$ we obtain

\begin{equation}
\Gamma\left(h,\sigma,\zeta,\lambda,\xi\right)\simeq\tilde{\Gamma}\left(h=0,\sigma,\lambda\right)e^{\sigma}\left(1+\frac{\sigma\zeta}{4}+\frac{\xi^{2}}{6}\mathcal{S}\right)\label{eq:RRExpanded}
\end{equation}
where $\mathcal{S}\left(\lambda\right)$ is defined in Eq. (\ref{eq:F_of_lambda}).
Note that $\tilde{\Gamma}\left(h=0,\sigma,\lambda\right)$ is given
in the second line of Eq. (6) in Ref. \onlinecite{dejardinetal08jpd}. 

Now, an expansion of Eq. (\ref{eq:VF}) with respect to $\theta_{\mathrm{VF}}/T$
yields\citep{landi13jap}

\[
\Gamma=\tau_{0}^{-1}\, e^{\frac{\Delta E}{k_{B}\left(T-\theta_{\mathrm{VF}}\right)}}\simeq\tau_{0}^{-1}\, e^{\sigma}\left(1+\sigma\frac{\theta_{\mathrm{VF}}}{T}\right)
\]
which is of the same form as our expression (\ref{eq:RRExpanded}).
Next, using Néel's approximation with a constant prefactor $\tau_{0}^{-1}$,
thus ignoring any dependence on temperature, damping and applied field,
$\tilde{\Gamma}\left(h=0,\sigma,\lambda\right)$ can be identified
with $\tau_{0}^{-1}$. Then, we can further identify the terms between
parentheses leading to the following expression for $\theta_{\mathrm{VF}}$
(in Néel's approximation)
\begin{equation}
\frac{\theta_{\mathrm{VF}}}{T}=\frac{\zeta}{4}+\frac{1}{6\sigma}\left(\xi^{2}\mathcal{S}\right).\label{eq:ThetaVFCorrected}
\end{equation}

\begin{figure}
\begin{centering}
\includegraphics[height=6cm]{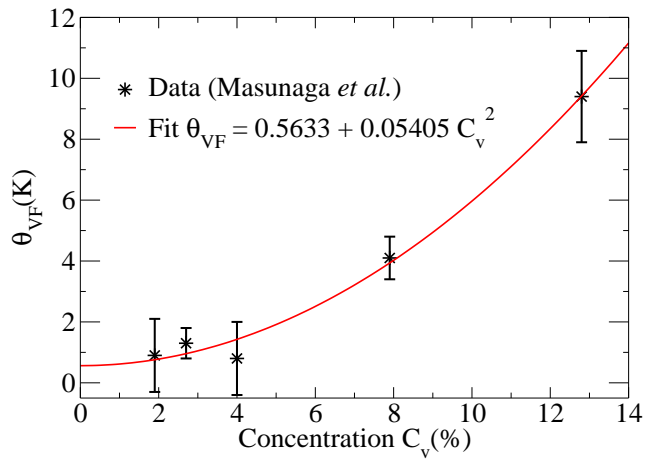}
\par\end{centering}

\caption{\label{fig:ThetaVFscaling}$\theta_{\mathrm{VF}}$ against the assembly
concentration. (stars) Experimental data from Masunaga \emph{et al.}\citep{masunagaetal09prb}
and (full line) fit of Eq. (\ref{eq:ThetaVFCorrected}).}
\end{figure}

This expression provides a somewhat ``microscopic'' description
of the phenomenological parameter $\theta_{VF}$ in terms of the inter-particle
interactions, the surface anisotropy and damping. Indeed, the last
term in (\ref{eq:ThetaVFCorrected}), which is similar to the one
derived in Ref. \onlinecite{landi13jap}, includes both the damping
parameter and the shape of the assembly, owing to the expression of
$\mathcal{S}\left(\lambda\right)$ {[}see Eq. (\ref{eq:F_of_lambda}){]}.
In addition, we note that $\xi$ is proportional to the assembly concentration\citep{sabsabietal13prb}
$C_{V}$ and thereby to $a^{-3}$, $a$ being the inter-particle distance.
Therefore, we expect that in the absence of surface anisotropy, $\theta_{\mathrm{VF}}$
scales as $\theta_{\mathrm{VF}}\sim C_{V}^{2}\sim a^{-6}$. In Ref.
\onlinecite{masunagaetal09prb} experimental estimates of $\theta_{\mathrm{VF}}$
are given for an assembly of Ni nanoparticles with varying concentration.
A comparison of Eq. (\ref{eq:ThetaVFCorrected}) with the corresponding
data is given in Fig. \ref{fig:ThetaVFscaling}.

On the other hand, the first term in Eq. (\ref{eq:ThetaVFCorrected})
accounts for the contribution from surface anisotropy. As discussed
earlier, in practice it should be possible to adjust the assembly
characteristics (assembly shape, particles size and underlying material)
so as to achieve to some extent a compensation between surface effects
and the DDI contribution. This could in principle suppress the dependence
of $\theta_{\mathrm{VF}}$ on the assembly concentration. In addition,
the term in $\zeta$ can also be used to extract from the experimental
data an estimate of the surface anisotropy coefficient $\zeta$ by
reading off the intercept from the plot in Fig. \ref{fig:ThetaVFscaling}.

In the most often encountered situation where the particles anisotropy
is modeled with an effective uniaxial anisotropy of constant $K_{\mathrm{eff}}$,
as would apply for elongated particles, dropping the $\zeta$ term,
the effective temperature $\theta_{\mathrm{VF}}$ explicitly reads
(as a function of the assembly concentration $C_{V}$)
\begin{eqnarray}
k_{{\rm B}}\theta_{\mathrm{VF}} & = & \left(\frac{\mu_{0}}{4\pi}\right)^{2}\frac{\left(M_{s}^{2}V\right)^{2}}{K_{\mathrm{eff}}V}\frac{\mathcal{S}}{6}\times C_{V}^{2}.\label{eq:TvfCv}
\end{eqnarray}

For example, consider a monodisperse assembly of spherical cobalt
nanoparticles of $3$ nm in diameter with $M_{s}\simeq1.4\times10^{6}\,{\rm J.T^{-1}.m^{-3}}$,
$K_{{\rm eff}}\simeq5\times10^{5}\,{\rm J.m^{-3}}$ , and $C_{V}\simeq1\%$.
Then, if the assembly is assumed to be in the form of a box-shaped
sample with its particles arranged into a simple cubic lattice, the
lattice sums $\mathcal{R}$ and $\mathcal{T}$ were given earlier
in the thermodynamic limit. Then, using $F\left(\lambda\right)\simeq1$
and $S_{2}\simeq1$, the factor $\mathcal{S}$ evaluates to $\mathcal{S}\simeq45$.
This yields $\theta_{\mathrm{VF}}\simeq0.05\,{\rm K}$, which is small
compared to the particle's blocking temperature $T_{{\rm B}}\simeq14\,{\rm K}$.
However, one should keep in mind that $\theta_{\mathrm{VF}}$ scales
with the particle's volume. 

It is worth emphasizing the fact that $\theta_{\mathrm{VF}}$ given
by Eq. (\ref{eq:TvfCv}) is independent of temperature, as can be
often encountered in the literature. However, if we take account of
surface anisotropy, Eq. (\ref{eq:ThetaVFCorrected}) shows that the
phenomenological parameter $\theta_{\mathrm{VF}}$ is in fact a linear
function of temperature via the term in $\zeta$. This can be understood
by noting that surface anisotropy, which is of cubic nature in the
EOPS model, drastically modifies the energy potential and thereby
affects the dynamics of the particle's magnetization. As a consequence,
the effect of DDI becomes strongly dependent on the thermal fluctuations
and the elementary switching processes they induce.

\section{Conclusion}

We have studied the combined effects of surface anisotropy and dipolar
inter-cluster interactions on the dynamic response of a mono-disperse
assembly of magnetic nanoclusters with textured anisotropy. We have
derived semi-analytical expressions for the in-phase and out-of-phase
components of the AC susceptibility as functions of temperature, applied
field, surface anisotropy, damping, frequency, and (weak) dipolar
interactions. If we ignore the surface anisotropy, we recover the
well known results of frequency- and interaction-induced shift in
both the maximum of $\chi^{\prime}$ and $\chi^{\prime\prime}$ and
of the temperature $T_{\max}$ thereat, taking into account the effect
of the assembly shape (oblate or prolate). In the presence of surface
anisotropy we have derived and used a semi-analytical expression for
the relaxation time and investigated the effect of surface (cubic)
anisotropy. We have done so in the limit of small field, high uniaxial
anisotropy barrier and weak surface anisotropy. The expressions obtained
for the small $\zeta$ show that the relaxation rate or the switching
probability increases with surface anisotropy, but the equilibrium
susceptibility decreases, thus leading to an overall upward shift
of $T_{\max}$. When the inter-particle interactions are switched
on, a competition sets in between the latter and surface anisotropy
that may lead, in adequately prepared samples, to a mutual compensation
of the two effects. 

Finally, our results for the relaxation rate have been analyzed in
connection with the so-called Vogel-Fulcher law and an expression
for the \emph{ad hoc }effective temperature has been given in terms
of the inter-particle dipolar interactions, the intra-particle surface
anisotropy and the damping parameter, in addition to the other physical
parameters such as the applied magnetic field and uniaxial anisotropy.
\begin{acknowledgments}
F.V. and H.K. would like to thank Denis Ledue for discussions and
useful comments in the early stage of this work.
\end{acknowledgments}
\bibliography{hkbib}

\end{document}